\documentclass[10pt]{iopart}

\usepackage{soul,color}
\usepackage{iopams}
\usepackage{setspace}
\usepackage{graphicx} 
\usepackage{capt-of} 
\usepackage{amstext}
\usepackage[ngerman,english]{babel}
\usepackage{xcolor}
\usepackage{booktabs}
\usepackage{cite}
\usepackage{hyperref} 
\hypersetup{colorlinks=false, citebordercolor=green}
\definecolor{colorThree}{rgb}{0.61, 0.87, 1.0}
\definecolor{colorTwo}{rgb}{0.99, 0.93, 0.0} 
\definecolor{colorOne}{rgb}{0.5, 1.0, 0.0}

\begin{document}
\title{Correlation induced localization of lattice trapped bosons coupled to a Bose-Einstein condensate}

\author{Kevin Keiler $^1$, Sven Kr\"onke $^{1,2}$ and Peter Schmelcher $^{1,2}$}
\address{$^1$Zentrum f\"ur Optische Quantentechnologien, Universit\"at
	Hamburg, Luruper Chaussee 149, 22761 Hamburg, Germany}
\address{$^2$The Hamburg Centre for Ultrafast Imaging, Universit\"at
	Hamburg, Luruper Chaussee 149, 22761 Hamburg, Germany}
\ead{\mailto{kkeiler@physnet.uni-hamburg.de}, \mailto{pschmelc@physnet.uni-hamburg.de}}

\onehalfspacing

\begin{abstract}
	We investigate the ground state properties of a lattice trapped bosonic system coupled to a Lieb-Liniger type gas. Our main goal is the description and in depth exploration and analysis of the two-species many-body quantum system including all relevant correlations beyond the standard mean-field approach. To achieve this, we use the Multi-Configuration Time-Dependent Hartree method for Mixtures (ML-MCTDHX). Increasing the lattice depth and the interspecies interaction strength, the wave function undergoes a transition from an uncorrelated to a highly correlated state, which manifests itself in the localization of the lattice atoms in the latter regime. For small interspecies couplings, we identify the process responsible for this cross-over in a single-particle-like picture. Moreover, we give a full characterization of the wave function's structure in both regimes, using Bloch and Wannier states of the lowest band, and we find an order parameter, which can be exploited as a corresponding experimental signature. To deepen the understanding, we use an effective Hamiltonian approach, which introduces an induced interaction and is valid for small interspecies interaction. We finally compare the ansatz of the effective Hamiltonian with the results of the ML-MCTDHX simulations.
\end{abstract}


\section{Introduction}
\sethlcolor{colorThree}
Nowadays, ultracold gases stand out due to the high degree of controllability, especially of trapping potentials and inter-atomic interactions. Thereby, capturing the atoms of a gas by using light fields allows for the realization of a variety of even rather complex many-body systems.
Especially one-dimensional (1D) systems are under intense investigation and show unique properties \cite{1d_1,1d_2}. Here, the inverse scaling of the effective interaction strength to the density allows for entering the strongly correlated regime in the dilute regime \cite{1d_eff_corr_1,1d_eff_corr_2}.
Moreover, the transversal confinement allows for a tuning of the interaction strength among the atoms via a so-called confinement induced resonance \cite{confine}. Alternatively, the same effect could be achieved via an atom-molecule Feshbach resonance \cite{feshbach}. This freedom of adjusting the interaction strength makes it feasible to enhance the deviation from the mean-field behaviour of the bosons, in a very controlled and systematic manner. Therefore, one-dimensional ultracold gases are particularly suited to access physical regimes, in which effective free theories cease to be valid.
Such strongly interacting Bose gases tend to be very sensitive to additional external trapping potentials such as lattices \cite{trans_opt_lat,TG_lat}. Here, a Mott insulating state can be formed for arbitrarily weak lattice amplitudes, in contrast to the conventional superfluid to Mott insulator transition \cite{greiner,jaksch}. \par
Correlations can not only appear within one type of species, but in particular between different bosonic species \cite{spec_el,wiemanRB}, offering a rich phenomenology. Due to the fact that we add to the already present intra-species interactions in the respective subsystems an interspecies interaction, mixtures show a plethora of intriguing phenomena, such as pair-tunneling effects \cite{pflanzer1,pflanzer2} and paired superfluidity \cite{pair_super1,pair_super2}.
Setups of impurities in a bath can be viewed as such a species mixture, covering the aspects mentioned above, and are a subject of ongoing research. They have been studied in various cases, e.g. the transport and the related collisions of the impurity through the bath \cite{trans_imp} and correlation effects due to the entanglement of the species \cite{knoerzer}. In particular, one-dimensional impurity-bath systems exhibit large interaction effects, bringing to light many peculiar phenomena \cite{gangardt,schecter_njp,schecter_annphys,zvonarev,naegerl_bloch}.
In addition, impurities in Bose-Einstein condensates can be exploited as a quantum simulator for polaron physics \cite{blume_imp,cuc_imp,corn_pol_sim}.
One of the first theoretical descriptions of polarons, including phonon clouds, was introduced by Fr{\"o}hlich \cite{froel}. Since then a lot of progress has been made in the limiting cases of weak \cite{llp,schultz} and strong electron-phonon coupling \cite{landau}.
While the ongoing theoretical study of one-dimensional polarons \cite{massignan_review,grusdt,volosniev,zinner_pol,garcia} has predicted a lot of intriguing properties, recent experiments \cite{trans_imp,naegerl_bloch,catani,fukuhura} have finally opened the door to the implementation of one-dimensional polaronic systems, providing a deeper understanding of 1D polarons.
When immersing more than one impurity in the bath, an induced interaction among the polarons appears, which counteracts the repulsive interaction among the impurities. The description in terms of polarons is in general of major interest for the understanding of the electron-phonon coupling in condensed matter physics.
In order to manipulate impurities in a controlled, systematic way in ultracold physics, it would be useful to load the impurities first in a lattice, which is in turn inserted into the bath, since lattices allow for single-site excitation as well as collective excitations. 
To some extent, such a setup has been investigated in the tight-binding limit recently  \cite{jaksch_pol,jaksch_clus,jaksch_trans}, where the authors especially focussed on the behaviour of the combined systems under the influence of increasing temperature, e.g. the clustering of polarons in the wells of the lattice due to an attractive induced interaction in dependence of the temperature. \par
In the present work, we focus on the role of interspecies correlations of lattice trapped atoms immersed in a Lieb-Liniger gas, and in particular how it impacts the structure of the many-body wave function. 
After a brief description of the specific system under investigation, we present the phenomenology of the ground state as a function of the lattice trap depth and interspecies interaction strength. Via the introduction of a correlation measure, we can identify a cross-over diagram exhibiting a transition from an uncorrelated to a strongly correlated state. We unravel the nature of the cross-over in an effective single-particle picture for small interspecies couplings. Moreover, we give a full characterization of the wave function in the limiting cases of weak and strong correlations. This enables us to derive and understand the properties of the ground state, employing the structural form for the ground state wave function. Finally, we aim at describing the correlated state using an effective Hamiltonian approach  \cite{nakajima}, thereby introducing an induced attractive interaction between the atoms trapped by the lattice, as well as an induced hopping. We find qualitative agreement with the full ML-MCTDHX calculations and identify the weak interspecies interaction regime, to which this effective approach is applicable.
\sethlcolor{colorTwo}
\section{Setup, Hamiltonian and methodology}
Our system consists of bosons trapped in a one-dimensional lattice with periodic boundary conditions, which is immersed in a Lieb-Liniger-like gas \cite{LL1,LL2,LLnagerl} of a second species of bosons. We note that this setup lies within reach of current experimental techniques, since beyond controlling the dimensionality, various trapping potentials for the atoms can be achieved \cite{boshier}, including in particular one-dimensional ring geometries. Moreover, it is possible to create an optical lattice potential, which does not affect the Lieb-Linger gas by choosing the right laser wavelengths and atomic species \cite{spec_sel_lat}. This allows for the creation of a two-component system with each species trapped individually on the same ring geometry. In the following, the species trapped by a lattice potential will be denoted as the A species, whereas the Lieb-Liniger-like gas refers to the B species. Furthermore, we introduce a coupling Hamiltonian $\hat{H}_{AB}$ between the two species. Both subsystems are confined to a longitudinal direction, accounting for the one-dimensional character, and excitations in the corresponding transversal direction can be neglected. This finally results in a Hamiltonian of the form $\hat{H}=\hat{H}_A+\hat{H}_B+\hat{H}_{AB}$. The Hamiltonian of the A species reads

\begin{equation}
\hat{H}_A=\int_{0}^{L} \text{dx} \; \hat{\chi}^{\dagger}(\text{x}) \Big [ -\frac{\hbar^{2}}{2 m_A} \frac{\text{d}^{2}}{\text{dx}^{2}} + V_0 \sin^{2}(\pi k \text{x}/L) \Big ] \hat{\chi}(\text{x}),
\end{equation}
where $\hat{\chi}^{\dagger}$ is the field operator of the lattice bosons, $m_A$ their mass, $k$ the number of wells in the lattice and $L$ is the circumference of the ring.
We focus on the regime, where interactions among the lattice atoms can be neglected, setting $g_{AA}=0$. The B species is described by the Hamiltonian of the Lieb-Liniger model

\begin{equation}
\hat{H}_B=\int_{0}^{L} \text{dx} \; \hat{\phi}^{\dagger}(\text{x}) \Big [ -\frac{\hbar^{2}}{2 m_B} \frac{\text{d}^{2}}{\text{dx}^{2}} + g_{BB} \; \hat{\phi}^{\dagger}(\text{x})\hat{\phi}(\text{x}) \Big ] \hat{\phi}(\text{x}),
\end{equation}
where $\hat{\phi}^{\dagger}$ describes the field operator of the Lieb-Liniger gas atoms, $g_{BB}>0$ is the interaction strength of the two-body contact interaction among the B species and $m_B$ the mass of the B species atoms. Moreover, we assume equal masses for the species $m_A=m_B$. The coupling between the species is given by

\begin{equation}
\hat{H}_{AB}= g_{AB} \int_{0}^{L} \text{dx} \; \hat{\chi}^{\dagger}(\text{x}) \hat{\chi}(\text{x}) \hat{\phi}^{\dagger}(\text{x}) \hat{\phi}(\text{x}).
\end{equation}
Throughout this work we consider a triple-well and we focus on the scenario of small particle numbers with three lattice atoms $N_A=3$ and ten atoms in the Lieb-Liniger gas $N_B=10$. The interaction among the latter atoms is set to a value where the depletion is negligible in case of no interspecies coupling, i.e. $g_{BB}/E_R \lambda=6.8 \times 10^{-3}$, with $E_R=(2\pi\hbar)^{2}/2m_A \lambda^{2}$ being the recoil energy and $ \lambda=2L/k$ the optical lattice wavelength. Therefore, we view the Lieb-Liniger gas as a Bose-Einstein condensate (BEC), in which impurities of species A are immersed. In particular, we shall analyze the ground state of the coupled system for different values of the repulsive interspecies interaction strength $g_{AB}$ and the lattice depth $V_0$. \par
Our numerical simulations are performed with the \textit{ab-initio} Multi-Configuration Time-Dependent Hartree method for bosonic (fermionic) Mixtures (ML-MCTDHX) \cite{mlb1,mlb2,mlx}, which takes all correlations into account and therefore allows for going beyond the lowest-band and tight-binding approximation for the lattice system and beyond the Bogoliubov approximation for the BEC. Within ML-MCTDHX one has access to the complete many-body wave function which allows us consequently to derive all relevant characteristics of the underlying system. Besides investigating the quantum dynamics it allows us to calculate the ground (or excited) states. Compared to the standard approach for solving the time-dependent Schr{\"o}dinger equation, where one constructs the wave function as a superposition of time-independent basis states with time-dependent coefficients, the ML-MCTDHX approach considers a co-moving time-dependent basis on different layers in addition to time-dependent coefficients. This leads to a significantly smaller amount of basis states that are needed to obtain an accurate description and eventually reduces the computation time. Moreover, the multi-layering of the method allows for a construction of the total wave function $|\Psi\rangle$, as sum of species product states, using the Schmidt decomposition \cite{schmidt}

\begin{equation}
	|\Psi \rangle=\sum_{i} \sqrt{\lambda_i} |\Psi^i_A \rangle \otimes |\Psi^i_B \rangle.
	\label{schmidt_dec}
\end{equation}
With the aid of this wave function decomposition, we are able to characterize and introduce interspecies correlations in a controlled and systematic manner. The case of a single contributing product state in the sum we call the species mean-field case, whereas deviations from it indicate interspecies correlations. The reader should note that the species mean-field is not to be confused with the Gross-Pitaevskii mean-field, which restricts the orbitals in each subsystem to a single one. In contrast to that, subsystems in the species mean-field case are allowed to carry arbitrary correlations and may be described by many contributing (optimized) orbitals.

\section{Ground state transition}
\begin{figure*}[t]
	\centering
	\includegraphics[scale=0.11]{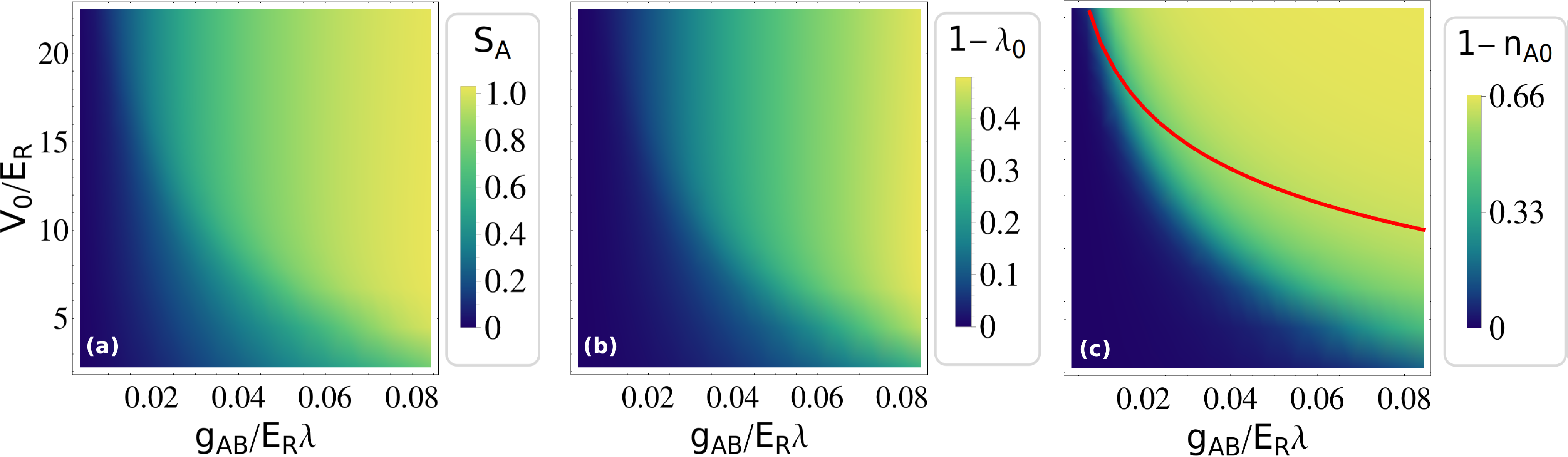}
	\captionof{figure}{(a) Von Neumann entropy $S_A$, (b) species depletion $1-\lambda_0$ and (c) depletion $1-n_{A0}$ of the most populated natural orbital of species A as a function of the lattice depth $V_0$ and the interspecies interaction strength $g_{AB}$. The species depletion is defined as the depletion of the most populated product state, when constructing the system's wave function using the Schmidt decomposition (Eq. \ref{schmidt_dec}). The red line in (c) refers to the single-particle argumentation in section \ref{sec_Bloch}.}
	\label{fig1_new}	
\end{figure*}

In the following, we investigate ground state properties of our system for different values of the lattice depth and the interspecies interaction strength, strongly focussing on intra- and interspecies correlations, that appear beyond mean-field. Thereby, we view the A species as an impurity species which is immersed into a homogeneous background gas. In the case of $g_{AB}=0$ both subsystems can be described separately and no interspecies correlations appear in the system's ground state. For the lattice atoms, this means that each atom occupies the energetically lowest Bloch state, and is therefore delocalized, since $g_{AA}=0$. Increasing the coupling strength $g_{AB} > 0$, disturbs this uncorrelated state. As a measure for correlations, we consider the eigenvalues $n_{Ai}$ and eigenvectors $|n_{Ai}\rangle$ of the one-body density operator $\hat{\rho}^{(1)}_A$ of the A species \footnote{We are well aware of the fact that fragmentation might also occur in the case of a ground state wave function, which can be represented by a single product state in the species mean-field. In order to be precise, one has to take additionally the species depletion or von Neumann entropy into account.}, i.e. the natural populations and natural orbitals \cite{natorb1,natorb2,natorb3}.
The one body-density operator is defined as the partial trace of the density operator $\hat{\rho}=|\Psi\rangle \langle \Psi|$, $|\Psi\rangle$ being the ground state wave function, over all particles except for one particle of the A species
\begin{equation}
\hat{\rho}^{(1)}_A=\Tr_{N_B, \; N_A-1}[\;\hat{\rho}\;]=\sum_{i} n_{Ai} |n_{Ai}\rangle\langle n_{Ai}|,
\end{equation}
where $\sum_{i}n_{Ai}=1$. In the case of a single non-zero eigenvalue equalling unity ($n_{A0}=1$), the particles in subsystem A can be considered as uncorrelated, while deviations from this value indicate correlations. In Fig. \ref{fig1_new} (c), we plot the depletion $1-n_{A0}$ of the most populated natural orbital. 
As can be seen for small values of $V_0$ and $g_{AB}$, the depletion is approximately zero, describing a state which shows no significant correlations for the atoms of species A. Increasing the lattice depth or interspecies interaction strength leads to a stronger deviation from the mean-field behaviour. The greenish area indicates the transition between an uncorrelated and a correlated ground state. In the yellow region, a correlated state with a depletion of $1-n_{A0}=2/3$ is reached. Interestingly, we find here that only two other natural orbitals become significant, whose populations saturate for deep lattices and strong interspecies interaction to  $n_{A1}=n_{A2}=1/3=n_{A0}$, leading to a triple degeneracy.
In order to see that this effect is due to physics beyond the species mean-field picture, it is useful to analyze the von Neumann entropy
\begin{equation}
S_A=-\Tr[\,\hat{\rho}_A\ln(\hat{\rho}_A)\,] \quad \text{and} \quad \hat{\rho}_A=\Tr_{N_B}[\;\hat{\rho}\;]
\end{equation} 
for the subsystem A in the different regimes. Here, $\hat{\rho}_A$ is defined as the trace of the density operator over all particles of the B species. In case the A species is in a pure state, the entropy $S_A$ equals zero, whereas a mixed state will lead to deviations from zero.
In Fig. \ref{fig1_new} (a) and (b), it becomes clear that for high lattice depth and interspecies interaction strength the wave function of the coupled system can no longer be described by a product ansatz of the subsystems. Hence, a species mean-field description is not valid and one has to go beyond it. In the representation of the wave function by a sum of products of species functions (Eq. \ref{schmidt_dec}), two additional degenerate product states (w.r.t $\lambda_i$) become important. For large values of $V_0$ and $g_{AB}$ the interspecies correlations are dominant, such that the contribution of these product states is of the same order as the previously, i.e. for smaller values of $V_0$ and $g_{AB}$, dominant one, resulting in a strongly entangled state.
\begin{figure}[t]
	\centering
	\includegraphics[width=0.7\linewidth,height=0.55\linewidth]{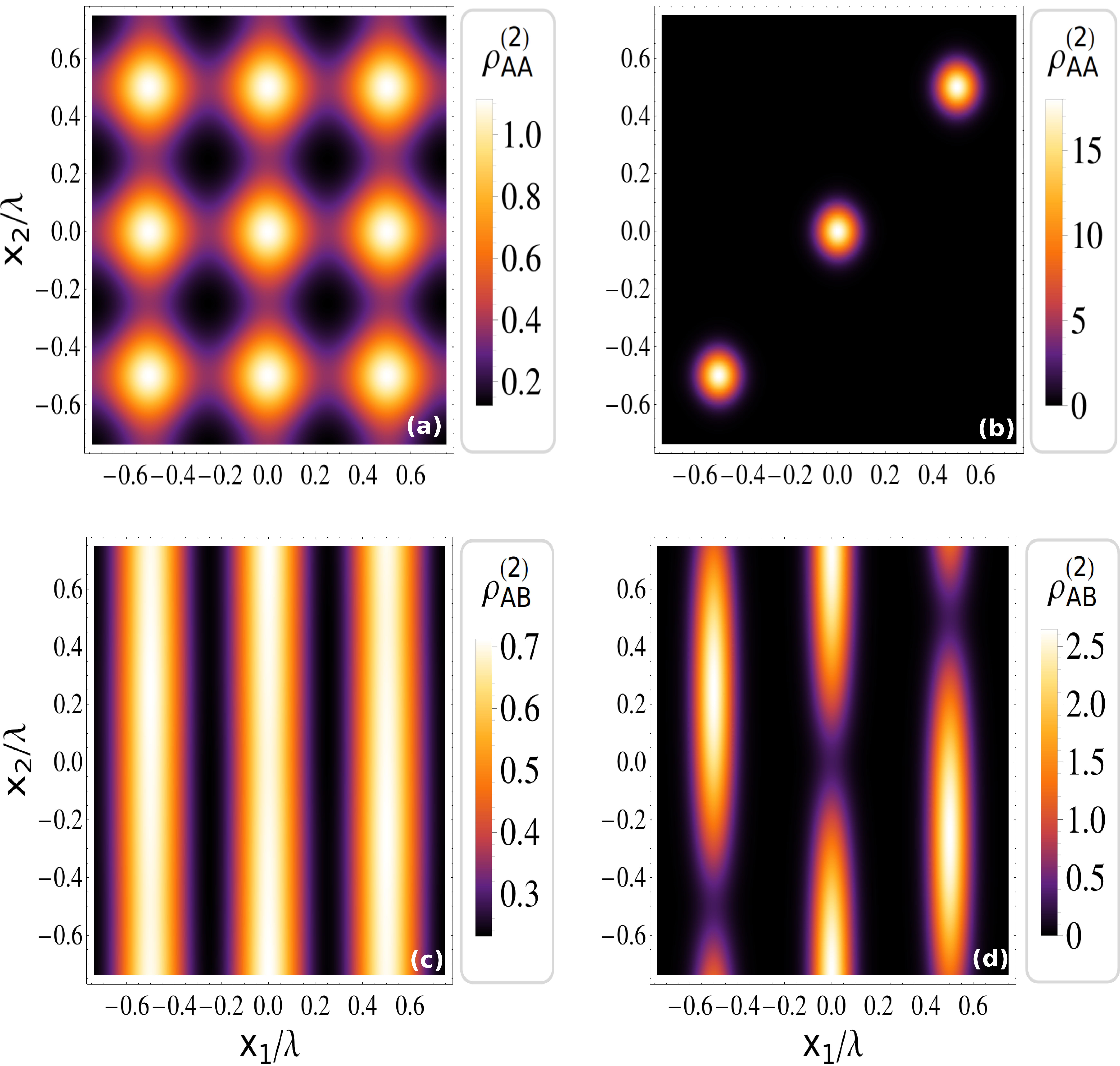}
	\captionof{figure}{Two-body density  $\rho^{(2)}_{AA}(x_1,x_2)$ (first row) and $\rho^{(2)}_{AB}(x_1,x_2)$ (second row) in the uncorrelated regime (a,c) with $g_{AB}/E_R \lambda=0.0034$ and $V_0/E_R=2.25$ and in the correlated regime (b,d) with $g_{AB}/E_R \lambda=0.084$ and $V_0/E_R=22.5$.}
	\label{fig_tbd}	
\end{figure} 
In order to understand the physical processes behind the transition from an uncorrelated to a correlated state, one naturally as a first step would investigate the spatial density profile, i.e. the one-body density, of the species A. While expecting a change in the spatial atomic density distribution with varying $V_0$, what we indeed find is a comparatively minor change of the density profile crossing the transition region of the cross-over diagram with only slightly enhanced localization in the wells due to a deeper lattice. 
Instead, the effect of correlations can be observed in the two-body density 
\begin{equation}
\rho^{(2)}_{AA}(\text{x}_1,\text{x}_2)=\Tr_{N_B, \; N_A-2}[\;\hat{\rho}\;],
\end{equation}
which describes the probability of finding one lattice atom at the position $\text{x}_1$ and another one at $\text{x}_2$. We note here, that all appearing densities are normalized to unity. In Fig. \ref{fig_tbd}, a drastic difference of the distribution of the two-body density $\rho^{(2)}_{AA}(\text{x}_1,\text{x}_2)$ for the two regimes, identified in Fig. \ref{fig1_new} (c), is visible. In the uncorrelated state, the existence of a particle at the position $\text{x}_1$ has no effect on the measurement probability of a second lattice atom at $\text{x}_2$, see Fig. \ref{fig_tbd} (a). Moreover, here one can see that both lattice atoms are delocalized over the lattice, as it is expected in case of almost decoupled subsystems A and B. For higher values of the lattice depth and the interspecies interaction strength, the effect of correlations becomes evident, see Fig. \ref{fig_tbd} (b). They manifest themselves in a localization of lattice atoms in the wells. In other words, a detection of an atom in any well is followed by a definite second measurement of another atom in that same well. Please note that this effect is not due to direct (attractive) interaction among the lattice atoms, since the latter is set to zero. Instead, here it is the interspecies interaction with the B species atoms, which induces these correlations to the A atoms. 
Moreover, as we have already discussed above in the context of Fig. \ref{fig1_new} (b), also interspecies correlations are present in the strongly correlated regime suggesting the two-body density
\begin{equation}
\rho^{(2)}_{AB}(\text{x}_1,\text{x}_2)=\Tr_{N_B-1, \; N_A-1}[\;\hat{\rho}\;],
\end{equation}
as a valuable observable. The latter describes the probability of finding one lattice atom at the position $\text{x}_1$ and a B atom at position $\text{x}_2$. In Fig. \ref{fig_tbd} (c), we find for an almost decoupled system that an initial measurement of a lattice atom has no visible impact on the distribution of a B atom. However, in the correlated regime [Fig. \ref{fig_tbd} (d)] the probability of measuring a B atom at the same position as the lattice atom drops to zero. In conclusion, the correlated state can be understood as a localization of A atoms while expelling the B atoms from the position of the latter. Since all wells are energetically equivalent, this holds for each well.

\section{Physical mechanisms and state characterization}
In this section, we explore the physical mechanisms underlying the ground state transition described in the previous section. Firstly, we focus on the regime of small interspecies interaction strengths, thereby comparing the interspecies interaction energy with the width of the first Bloch band. We provide then a full characterization of the system's wave function in terms Wannier states and derive the relevant observables. Finally, we use the transformation of the Hamiltonian into the so-called Nakajima frame, in order to calculate the ground state in an effective model approach. This will help us to develop an understanding for the complex many-body physics in a simplified picture, based on induced interactions.
\subsection{Mechanism of the transition}\label{sec_Bloch}
As stated above, the localization of the lattice atoms of the A species, reminiscent of the phase separation of two Bose gases \cite{phase_sep}, is solely induced by the interaction with the BEC of B atoms. Therefore, we should observe a change of the interspecies interaction energy, which is defined as the expectation value $\langle \hat{H}_{AB} \rangle$ with respect to the ground state, for increasing lattice depth and interspecies interaction strength.
\begin{figure}[b]
	\centering
	\includegraphics[scale=0.07]{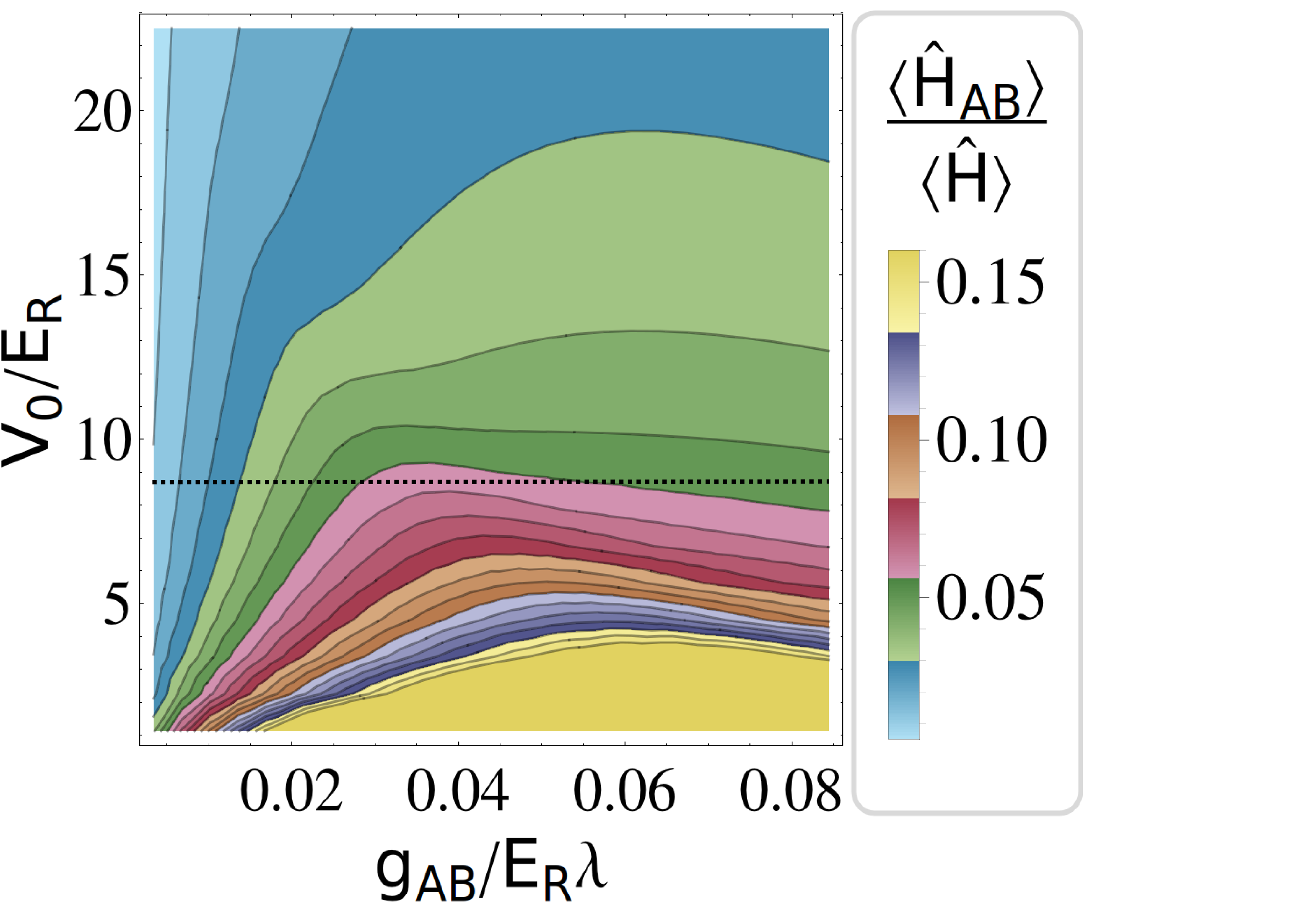}
	\caption{Interspecies interaction energy $\langle \hat{H}_{AB}\rangle$ in relation to the total energy $\langle \hat{H}\rangle$. The black horizontal line indicates, that for a fixed lattice depth the interspecies energy increases with the interspecies interaction strength $g_{AB}$ up to a certain value. Thereafter, $\langle \hat{H}_{AB}\rangle$ decreases monotonically.}
	\label{intE}	
\end{figure}
For a fixed lattice depth, it turns out that instead of monotonously increasing with the interaction strength $g_{AB}$, the interspecies energy reaches a maximum at a certain value of $g_{AB}$ and decreases for even higher values of $g_{AB}$ (Fig. \ref{intE}). Interestingly, the points of maximal $\langle H_{AB}\rangle$ in the cross-over diagram appear in the transition region where correlations become important [cf. Fig. \ref{fig1_new} (c)]. Hence, the formation of correlations is connected to a reduction of the interspecies energy. The eigenstates of a single atom in a periodic potential are the Bloch states \cite{bloch}, which can be grouped into bands. For the case of three lattice sites with periodic boundary conditions, the lowest band consists of an energetically lowest Bloch state and two degenerate excited states. In the previous section, we have seen that in the uncorrelated regime ($g_{AB}\approx0$) the atoms of the A species are delocalized over the lattice, which is due the fact that each of them occupies exactly the energetically lowest Bloch state for $g_{AB}=0$. In the correlated regime, the ground state exhibits a localization of the A atoms. We can now interpret this localization as a superposition of the Bloch states of the first band, resulting, as a matter of fact, in principal in a Wannier state.
Therefore, we now aim at an explanation for the responsible mechanism in a single-particle-like picture. Based on the findings above, we assume that the reduction of the interspecies energy is related to a coupling of the Bloch states in the first band. Without any correlations between the species, using a product ansatz $|\Psi_\Pi\rangle=|\Psi_A\rangle \otimes|\Psi_B\rangle$, the interspecies energy is determined by the integral over the product of one-body densities of each subsystem
\begin{equation}
\langle\Psi_\Pi|\hat{H}_{AB}|\Psi_\Pi\rangle= g_{AB} N_A N_B \int_{0}^{L} \text{dx} \; \rho^{(1)}_A(\text{x}) \; \rho^{(1)}_B(\text{x}).
\end{equation}
This ansatz works well in the uncorrelated regime, where almost no interspecies correlations are present.
We observe that the one-body density of the B species exhibits rather small modulations, compared to the modulations of the one-body density of the A species (not shown here). The latter is due to the fact that $\rho^{(1)}_B(\text{x})$ is approximately homogeneous, which further approximates the interspecies energy to $\langle\Psi_\Pi|\hat{H}_{AB}|\Psi_\Pi\rangle=g_{AB}\, N_A  \,N_B/L$.
In a single-particle picture, we now argue that this interspecies energy per particle (of the A and B species) $g_{AB}/L$ is used, in order to excite a lattice atom occupying the energetically lowest Bloch state to one of the excited states in the first band. We use this as an estimation for the transition region and therefore search for the interaction strength $g_{AB}$ that leads to an interspecies energy per particle, equalling the band width for a given lattice depth $V_0$, i.e. being sufficiently large for this excitation process. In Fig. \ref{fig1_new} (c), we find that for small values of $g_{AB}$ the resulting curve (red) describes the phase border appropriately. For larger values this curve fails to describe the transition region indicating a break-down of the above given simple single-particle picture.  
We have seen that we can explain the onset of the transition from the uncorrelated state towards the strongly correlated state in an correlation-free effective single-particle picture for small $g_{AB}$, where sufficient interaction energy is supplied to excite the A species atoms within the first Bloch band.

\subsection{State characterization}
The fact that the two-body density suggests a localization of A atoms, paired with the possibility of excited Bloch states, motivates the following procedure. In this sense, it is intuitive to approximately describe the ground state in the correlated regime in terms of number states, spanned by the Wannier states of the lowest band. The Wannier states we employ are generalized ones, i.e. the eigenstates of the position operator projected onto the respective band \cite{kivelson1,kivelson2}. In order to find out, whether the localization in Fig. \ref{fig_tbd} can be understood in terms of a number state with all particles residing in the same Wannier state for the entire correlated regime, we calculate the projection of the ground state wave function $|\Psi\rangle$ on that number state of species A and sum over all configurations for the B species. The corresponding number state notation reads $|n_1,n_2,n_3\rangle_W$, where $n_i$ denotes the number of particles in the corresponding Wannier states sorted from the left to the right well. We proceed in the same manner for the case of number states, which are spanned by Bloch states in the lattice potential. The corresponding state reads $|n_1,n_2,n_3\rangle_{\text{Bl}}$, where $n_1$ refers to the energetically lowest Bloch state, while $n_2$ and $n_3$ refer to the two degenerate excited ones in the first Bloch band. The probability of finding the ground state in a number state $|\vec{n}^{A}\rangle$ of the subsystem A is then defined as
\begin{figure*}[t]
	\centering
	\includegraphics[scale=0.09]{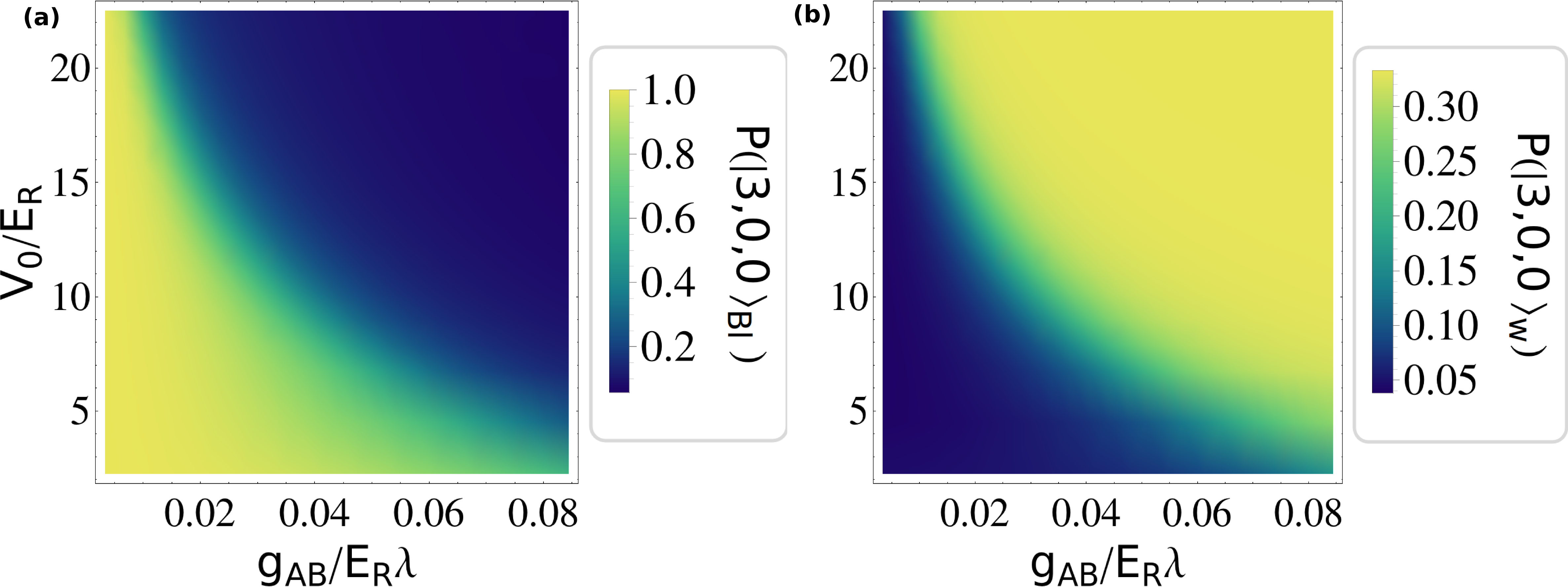}
	\caption{Probability of all three lattice atoms (a) residing in the energetically lowest Bloch state and (b) localized -exemplarily, since all wells are equally valid due to translational symmetry- in the left-most Wannier state of the lowest band.}
	\label{fig_prob}	
\end{figure*}
\begin{equation}
	P(|\vec{n}^{A}\rangle)=\sum_{i} |\langle \vec{n}^{i}_B|\otimes \langle \vec{n}^{A}|\Psi \rangle|^{2},
\end{equation}
where $\{|\vec{n}^{i}_B\rangle\}$ describes any number state basis set for the subsystem B with fixed particle number. For an uncorrelated state, we expect all the lattice atoms to reside in the energetically lowest Bloch state of the first band, since $g_{AA}=0$. In Fig. \ref{fig_prob} (a) we find, that in the uncorrelated regime the probability for all lattice atoms to be in the lowest Bloch state is $P(|3,0,0\rangle_\text{Bl})=1$. Compared to that, the probability for all particles to be e.g. in the left Wannier state is $P(|3,0,0\rangle_W)=\frac{1}{27}$, which can be derived analytically for $g_{AB}=0$. Increasing the values of the lattice depth and interspecies interaction strength, the energetically lowest Bloch number state gets depopulated. Instead, the lattice atoms strongly populate the Wannier number state, saturating towards a value for the probability of $P(|3,0,0\rangle_W)=\frac{1}{3}$. Due to translational symmetry this is true for all number states with all lattice atoms residing in the same well, meaning $P(|3,0,0\rangle_W)=P(|0,3,0\rangle_W)=P(|0,0,3\rangle_W)=\frac{1}{3}$. To summarize, for small values of $g_{AB}$ and $V_0$ the energetically lowest Bloch state is occupied by all A atoms, whereas for large values the correlated state is a superposition of number states, where all A atoms occupy the same Wannier state, implying their clustering in the corresponding wells.
Now that we know more about the structure of the system's wave function, we are able to specify the wave function in the corresponding regimes. In the uncorrelated regime, the ground state reads

\begin{equation}
|\Psi\rangle_\text{uc}\approx|3,0,0\rangle_{\text{Bl}}\otimes|\Psi_B\rangle,
\end{equation}
where $|\Psi_B\rangle$ is the ground state of the Hamiltonian $\hat{H}_B$. Based on the projection of the ground state for the correlated regime onto the respective number states [Fig. \ref{fig_prob} (b)], the wave function is of the form

\begin{equation}
|\Psi\rangle\approx\frac{1}{\sqrt{3}}\Big [|3,0,0\rangle_{W}\otimes|\Psi^{1}_B\rangle+|0,3,0\rangle_{W}\otimes|\Psi_B^{2}\rangle+|0,0,3\rangle_{W}\otimes|\Psi_B^{3}\rangle \Big],
\label{eq:wfn_corr}
\end{equation}
where $\{|\Psi^{i}_B \rangle\}$ are non-orthogonal basis states for the B species.
\begin{figure}[t]
	\centering
	\includegraphics[scale=0.2]{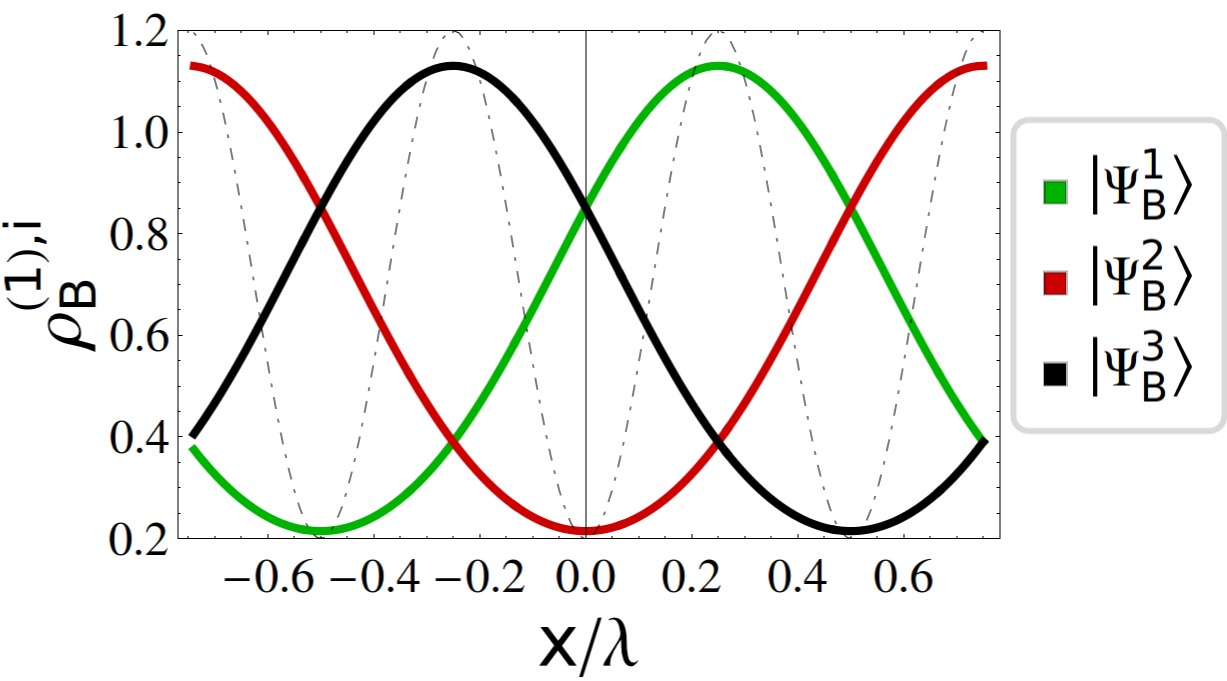}
	\captionof{figure}{One-body density $\rho^{(1),i}_B$ for the states of the BEC in Eq. \ref{eq:wfn_corr} with $g_{AB}/E_R \lambda=0.084$ and $V_0/E_R=22.5$. The dotted line is a sketch of the lattice potential, indicating the position of the wells.}
	\label{fig_dens_B}	
\end{figure}
Measuring a lattice atom in a specific well, we then know that all the other lattice atoms will be in that same well. Additionally, measuring the subsystem A to be in well $i$ implies a collapse of the ground state onto a definite wave function $|\Psi^{i}_B\rangle$ for the B species. Due to the repulsive interspecies interaction the density $\rho_B^{(1),i} = \Tr_{N_B-1}[\,|\Psi^i_B\rangle \langle\Psi^i_B| \,]$ exhibits a minimum at the well where the A atom was measured, shown in Fig. \ref{fig_dens_B}. 
Employing the structural form of the wave function given by Eq. \ref{eq:wfn_corr}, we can now derive the observed one- and two-body density of the A species atoms discussed above, using the Wannier states  $|w_i\rangle$, where the index $i$ corresponds to the localization in the $i$-th well. We obtain

\begin{equation}
\hat{\rho}^{(1)}_{A}=\frac{1}{3}\Big[\sum_{i=1}^{3}|w_{i}\rangle\langle w_{i}| \Big] \quad \text{and}
\label{eq_obdA_x}
\end{equation}

\begin{equation}
\rho^{(2)}_{AA}(x_1,x_2)=\frac{1}{3}\Big[\sum_{i=1}^{3}|w_{i}(x_1)|^2 \; |w_{i}(x_2)|^2 \Big].
\label{eq:tbdAA_x}
\end{equation}
Now, it is obvious that the natural populations of $\hat{\rho}^{(1)}_A$ are degenerate, equalling $\frac{1}{3}$, in the correlated regime.
The spatial structure of the two-body density $\rho^{(2)}_{AA}$ is governed by the overlap of the density of equal Wannier states in that regime. This corresponds to a two-dimensional representation of the Wannier states, so that one arrives at a localized structure in the two-body density [Fig. \ref{fig_tbd} (b)]. 

Above, we have identified the natural populations of the A species as a quantity, which reflects the transition from the correlated to the uncorrelated regime. While for our specific scenario it is sufficient for the characterization of the two regimes, other states, such as the Mott insulator state, cannot be distinguished from the correlated state. The latter shows the same behaviour in terms of natural populations as the A species in the correlated regime. Therefore, we cannot distinguish the Mott insulator state from the ansatz in Eq. \ref{eq:wfn_corr}, using only the natural populations.
Instead, it is useful to consider the variance of the particle number operator $\hat{n}_i=\hat{a}^{\dagger}_i\hat{a}_i$ for a specific site $i$, where $\hat{a}^{\dagger}_i$ is the creation operator for creating a lattice atom in the i-th Wannier state. Using the respective wave function ansatz, we can derive the variances $(\Delta n_i)^{2}=\langle\hat{n}^2_i \rangle - \langle \hat{n}_i\rangle^2$ for the different states. In table \ref{fig_var}, we see that the particle number variances differ strongly for the three states, allowing for a clear assignment and identification. Therefore, $(\Delta n_i)^{2}$ can be exploited as an experimental signature for the different states and thus also for correlations, since it is accessible via quantum gas microscopy \cite{microscope1,microscope2}.

\begin{table}[t]
	\centering
	\captionof{table}{Variance of the particle number operator $\hat{n}_i=\hat{a}^{\dagger}_i\hat{a}_i$ for a specific site $i$.}
	\label{fig_var}
	\begin{tabular}{cc}
		\br
		& $(\Delta n_i)^{2}$\\ \mr
		uncorrelated regime & $\frac{2}{3}$ \\ 
		correlated regime & $2$ \\ 
		Mott insulator & $0$ \\
		\br
	\end{tabular}
\end{table} 

\subsection{Induced interaction}
In subsection \ref{sec_Bloch}, we have described the transition region for small interspecies interaction strengths in terms of a coupling of Bloch states induced by the interspecies interaction. However, it is possible to gain a deeper understanding of the localization of A atoms, introducing an effective Hamiltonian for the A species which exhibits an induced interaction and an induced hopping term. With the aid of these, we will discuss the effective attractive interaction among the A species. The effective description in terms of an induced impurity-impurity interaction is a topic of ongoing research, ranging from Casimir-like forces between static impurities \cite{zwerger_casimir,fleischhauer_ind} to the inclusion of correlation effects for mobile impurities \cite{zinner_ind,jie_chen}. In order to arrive at an effective Hamiltonian, we need to apply the Nakajima transformation and use the Bogoliubov approximation for the B species, which finally decouples the subsystems \cite{froehlich,bardeen,bcs}. We note that due to the non-unitarity of the transformation, in general observables are not invariant under the transformation. Setting $\hbar=L=m_{A/B}=1$, the effective Hamiltonian $\hat{\bar{H}}$ in the transformed frame reads
\begin{equation}
\eqalign{\hat{\bar{H}}=\hat{\bar{H}}_0 + \hat{V}_\text{ind} \quad \text{where,}\\
	\hat{V}_\text{ind}=\frac{1}{2}\sum_{k,q,k^{\prime},q^{\prime}} U_{k,k^{\prime},q,q^{\prime}} \; \hat{a}^{\dagger}_k \hat{a}^{\dagger}_{k^{\prime}}\hat{a}_q\hat{a}_{q^{\prime}} \quad \text{and}\\
	\hat{\bar{H}}_0=\sum_{k} \epsilon_k \hat{a}^{\dagger}_k \hat{a}_k+\sum_{i} \omega_i \hat{b}^{\dagger}_i \hat{b}_i +\sum_{kq} t_{kq} \hat{a}^{\dagger}_k \hat{a}_q,
}
\label{eff_H}
\end{equation}
with $U_{k,k^{\prime},q,q^{\prime}}=\sum_{i} \Big[ \frac{g^{k^{\prime}q^{\prime}}_i (g^{qk}_i)^{\ast}}{\epsilon_k- \epsilon_q - \omega_i}- \frac{g^{kq}_i (g^{q^{\prime}k^{\prime}}_i)^{\ast}}{\epsilon_k- \epsilon_q + \omega_i} \Big]$ and $t_{kq}=\frac{1}{2} \sum_{r} U_{rkqr}$.
The derivation of $\hat{\bar{H}}$ can be found in the appendix.  $\hat{a}^{\dagger}_k$ is the creation operator for a particle in the Bloch state $|\chi_k \rangle$ with energy $\epsilon^{(1)}_k$, while  $\epsilon_k=\epsilon^{(1)}_k+g_{AB}$, and $\hat{b}^{\dagger}_i$ the one for the Bogoliubov mode $v^{\ast}_i$. The matrix elements $g^{kq}_i=\sqrt{N_B} g_{AB} \; \langle \chi_k| u_i(\text{x})+ v_i(\text{x})|\chi_q \rangle$ can be interpreted as a coupling of Bloch states, mediated by Bogoliubov modes. This coupling is resonantly enhanced, when the band width $\epsilon_k- \epsilon_q$, assuming $k\neq q$ and restricting to the lowest band, matches the energy of a Bogoliubov mode $\omega_i$.
On top of the induced interaction $\hat{V}_\text{ind}$, an additional one-body term, having the character of a hopping term with hopping elements $t_{kq}$, occurs in the Hamiltonian. The ground state of the total wave function is a product state of the ground state of the Bogoliubov Hamiltonian and the ground state of the residual Hamiltonian, consisting of the induced terms and $\hat{H}_A$. We can now easily calculate the ground state for the B species, which is given by the Bogoliubov vacuum. The calculation for the lattice A species is simplified, since there is no direct coupling to the B species. For all the upcoming considerations we restrict everything to the lowest band and set an upper-bound for the energy of the Bogoliubov modes. 
Before calculating the ground state for the A species, we analyze the tensor elements $U_{k,k^{\prime},q,q^{\prime}}$ and hopping elements  $t_{kq}$, thereby gaining an understanding of the induced attractive interaction.

\begin{figure}[t]
	\centering
	\includegraphics[width=0.6\linewidth,height=0.6\linewidth]{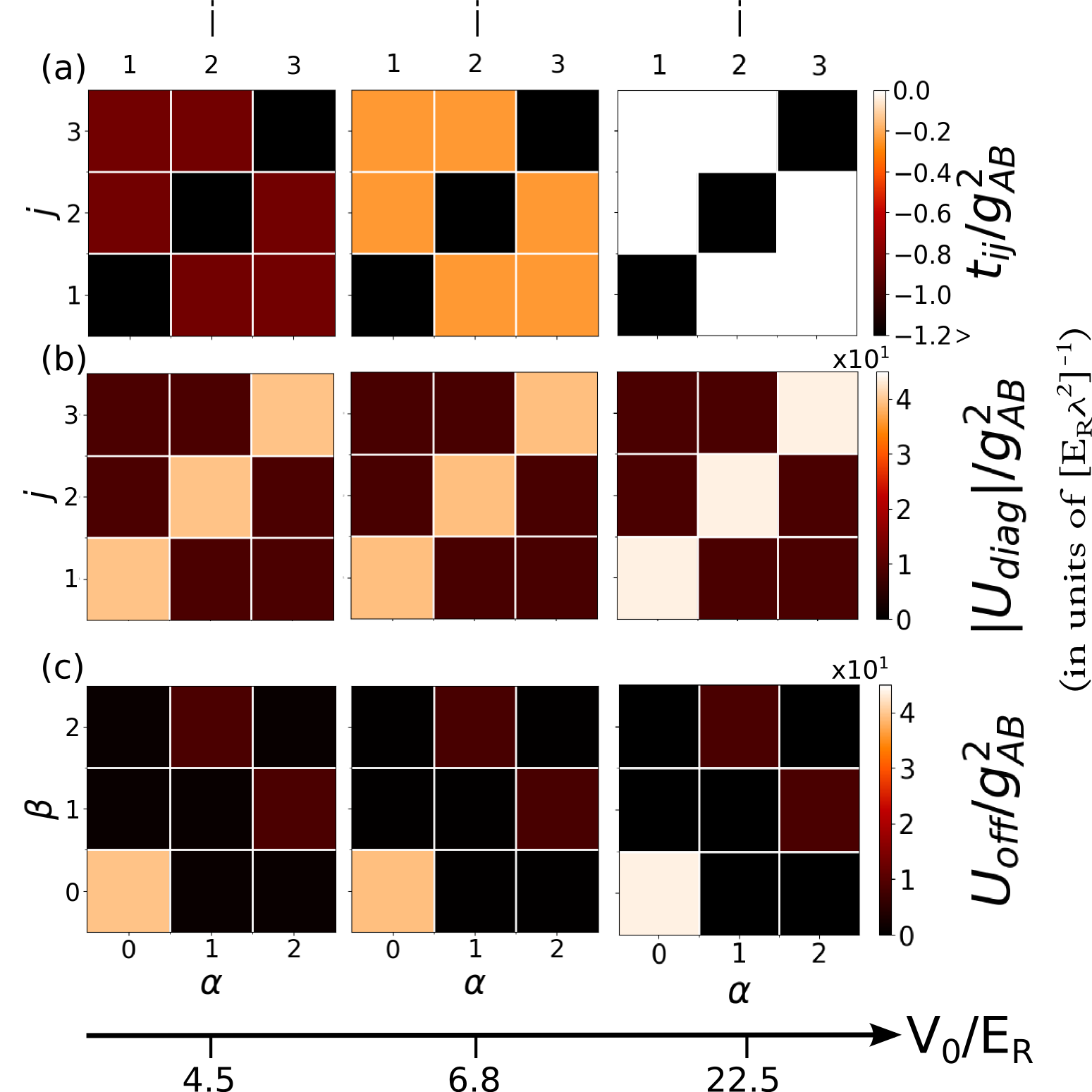}
	\caption{(a) Matrix elements of the one-body term $t_{ij}$, (b) diagonal of the tensor $U_{diag}=U_{i,j,i,j}$ and (c) a measure for the off-diagonal terms $U_{off}=\max_{i,j>0} |U_{i,j,i+\alpha,j+\beta}|$ (if e.g. $i+\alpha>3$, the resulting index is replaced by $[i+\alpha] \, \text{mod} \, 3$), deviating by $\alpha$ and $\beta$ from the diagonal elements, for different lattice depths $V_0$. The quantities are depicted in the Wannier basis, being restricted to the lowest band.}
	\label{mat}	
\end{figure}
In Fig. \ref{mat} we show the diagonal elements and a measure for the off-diagonal elements of the tensor $U$, as well as the induced hopping terms $t_{ij}$ for different values of the lattice depth.
In Fig. \ref{mat} (b), (c) we can see that the dominant elements are those for equal indices ($U_{i,i,i,i}$), followed by diagonal terms with respect to the particles ($U_{i,j,i,j}$). The particle exchange symmetry and the spatial translation symmetry is also reflected in the figures. All the other elements, which are off-diagonal, have almost no contribution compared to the diagonal terms. Moreover, for increasing $V_0$ the elements $U_{i,i,i,i}$, corresponding to the operator $\hat{a}^{\dagger}_i \hat{a}^{\dagger}_i \hat{a}_i\hat{a}_i=\hat{n}_i(\hat{n}_i-1)$, become larger. Since this element is negative and the most dominant one, the total term $U_{i,i,i,i}\hat{n}_i(\hat{n}_i-1)$ suggests an attractive on-site interaction among the A species. Still, we cannot make any definite statements concerning the structure of the wave function, based solely on the tensor elements $U_{i,j,k,l}$. Instead, it is useful to additionally analyze the behaviour of the matrix elements of the one-body term $t_{ij}$ for different lattice depths. Here, the off-diagonal terms are of interest, since they describe an induced hopping, whereas the diagonal elements just form an energetic offset. In Fig. \ref{mat} (a) we observe that the hopping tends towards zero for increasing $V_0$, which can be viewed as an indicator for the localization of the lattice atoms in this transformed frame. \par
\begin{figure}[t]
	\centering
	\includegraphics[width=0.7\linewidth,height=0.525\linewidth]{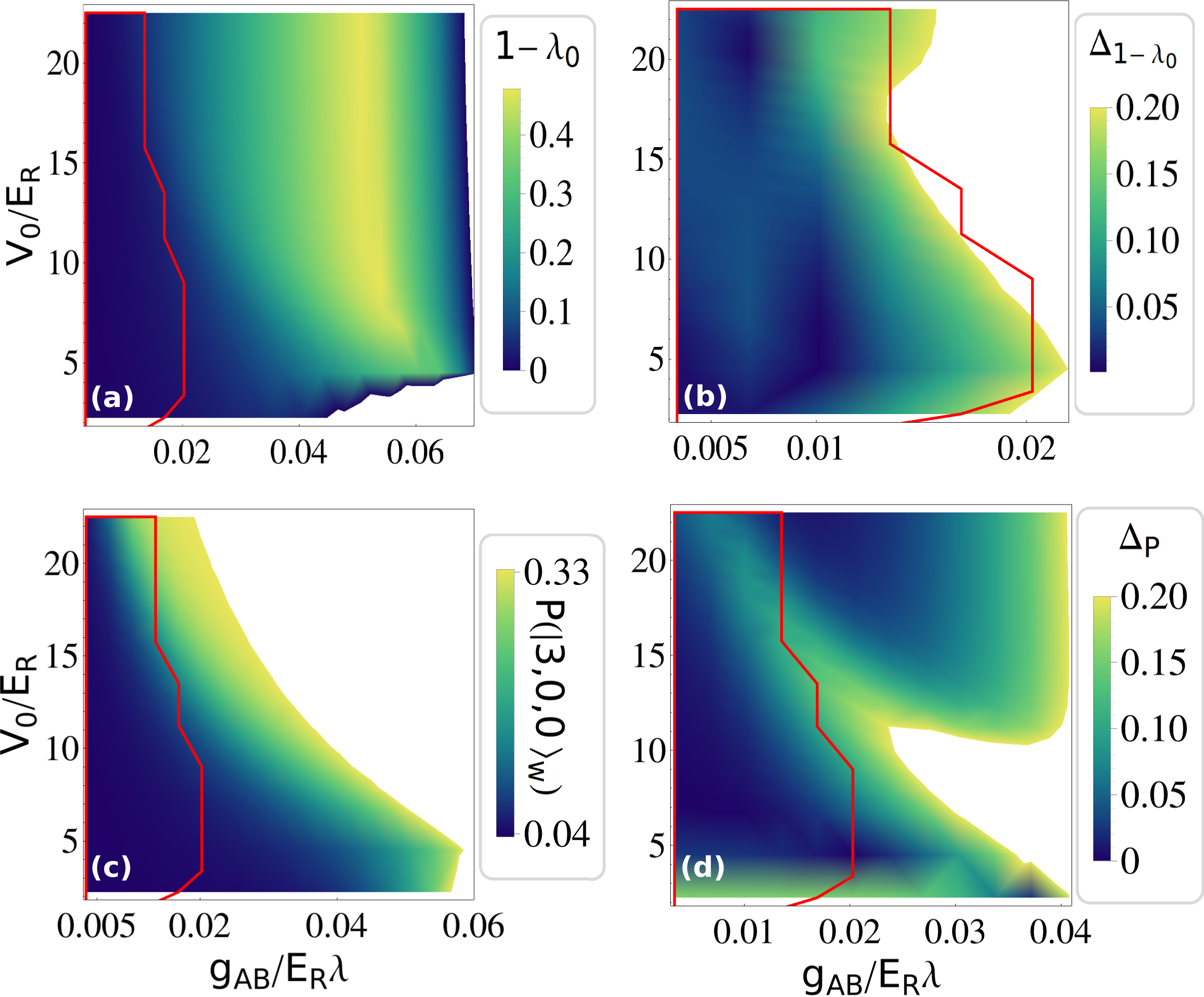}
	\caption{Results based on the effective theory (see Eq. \ref{eff_H}) and transforming back to the laboratory frame. (a) Species depletion [cf. Fig. \ref{fig1_new} (b)] and (b) the corresponding relative deviation from the ML-MCTDHX calculation $\Delta_{1-\lambda_0}=|[1-\lambda_0]_{\text{Nak}}-[1-\lambda_0]_{\text{M}} |/[1-\lambda_0]_{\text{M}}$. The index Nak refers to calculations using the Nakajima transformation, and M refers to the ML-MCTDHX simulations. (c) Probability of finding the system in the state $|3,0,0\rangle_W$ [cf. Fig. \ref{fig_prob} (b)] and (d) the relative deviation. The red boundary indicates the area of $g_{AB}$ and $V_0$ values for which the relative deviation of the total wave function's norm from unity is smaller than $0.3\%$. White spaces indicate that the effective theory is not valid (exhibiting non-physical values) in (a) and (c) or that the relative deviation is higher than $20\%$ in (b) and (d).}
	\label{rel_dev}	
\end{figure}
Now, one might expect that the ground state for the subsystem A will consist of localized lattice atoms for large lattice depths and interspecies interaction strengths. But since the above analysis is performed in the transformed frame, we finally have to transform back to the laboratory frame. In order to obtain the ground state of the coupled two-species system, we do the following: Firstly, we determine the ground state of the subsystem A, by performing a diagonalization of the Hamiltonian $\hat{\bar{H}}-\sum_{i} \omega_i \hat{b}^{\dagger}_i \hat{b}_i$ in the number state basis, spanned by the Wannier states of the lowest band. The ground state of the subsystem B is given by the Bogoliubov vaccum. The total ground state, consisting of a tensor product of both subsystem ground states, we transform back to the laboratory frame by applying the operator $\exp(-\hat{S})\approx 1-\hat{S}+\hat{S}^2/2$ (see appendix) to the total ground state up to second order. This is motivated by our initial assumption of small interactions, where terms of $\hat{S}$ which are higher than second order in $g_{AB}$ are neglected.
Finally, we compute the probability of finding the ground state in the number state $|3,0,0\rangle_W$, spanned by lowest band Wannier states, as well as the species depletion. Both quantities are available exactly within the ML-MCTDHX approach, being quantitatively well converged for all interspecies interaction strengths and lattice depths we investigated. For the calculations using the detour via the transformed frame, we expect appropriate results for small interspecies interaction strengths. Fig. \ref{rel_dev} shows the above-mentioned quantities and compares with the results of the ML-MCTDHX simulations in terms of a relative deviation.\par
At first glance, the species depletion in Fig. \ref{rel_dev} (a) resembles the one in Fig. \ref{fig1_new} (b) up to an interspecies interaction strength of $g_{AB}/E_R \lambda\approx0.05$. However, this resemblance is solely of qualitative nature, which becomes obvious in the relative deviation [Fig. \ref{rel_dev} (b)]. Here, we see a quantitative difference of the two methods, being of the order of up to $20\%$ for a maximum value of $g_{AB}/E_R \lambda\approx0.02$. A relative deviation of $20\%$ corresponds to a relative deviation of $|\langle\Psi|\Psi\rangle_{\text{Nak}}-1|\approx 0.3\%$ - due to the non-unitarity of the transformation - for the total wave function's norm from unity, where $|\Psi\rangle_{\text{Nak}}$ is the ground state wave function calculated by the effective approach. Also, the behaviour of the probability of finding the ground state in the number state $|3,0,0\rangle_W$ [Fig. \ref{rel_dev} (c)] qualitatively agrees with the results obtained by the ML-MCDTHX method, but again quantitatively deviates for large enough $g_{AB}$. But here, the range of validity w.r.t. the domains of $g_{AB}$ and $V_0$ values is more extended. Nevertheless, both the species depletion as well as the probability show the same qualitative behaviour we observe for the corresponding ML-MCTDHX results. In particular, with this effective Hamiltonian approach the cross-over diagram could be qualitatively reproduced for the uncorrelated regime, extending towards the border of the transition region and thereby indicating the latter. 
\section{Conclusions}
In the present paper we have investigated the appearance of correlations of a lattice trapped bosonic species coupled to a Bose-Einstein condensate. We have found a transition from an uncorrelated to a highly correlated state and characterized the associated quantum states from a many-body perspective. This transition is driven by interspecies correlations, bringing the system from an uncorrelated to a strongly correlated state. For small interspecies couplings we identified the transition region by energetic and mean-field arguments. Furthermore, we deduced expressions for the full system wave function for both regimes, using a Wannier and a Bloch basis analysis. In the uncorrelated state, the A atoms populate the lowest Bloch state and the B atoms are approximately homogeneously spread. In contrast, the correlated state is a superposition of states where all A atoms cluster together in this well, while the B atoms are expelled from it.
In order to measure the transition and in particular the correlated state, we propose to use the variance of the number of particles per well as an experimental signature and order parameter for correlations. Alternatively, we calculated the ground state, using the transformation into the Nakajima frame, which is valid for small interspecies  couplings. With this ansatz, we could reproduce only a small portion of the cross-over diagram for small lattice depths and interspecies interaction strengths, but gained a deeper insight into the role of induced interactions and induced hoppings for the process of localization. Eventually, this also demonstrated the uniqueness and power of the ML-MCTDHX method, which allowed - compared to the effective Hamiltonian ansatz - for calculations far beyond the mean-field approach. This analysis is also applicable for a larger number of particles in the environment, while still remaining in the few particle regime (we have tested this for $N_B\in[10,30]$), resulting in the same transition from an uncorrelated to a correlated state. However, such a particle increase will also increase the attractive induced interaction for a given choice of $V_0$ and $g_{AB}$, thereby shifting the transition region.  \par
The understanding of the cross-over in terms of a localized states analysis and induced interactions serves as a perfect starting-point for even more complex setups. For example, when introducing more lattice atoms or a repulsive intra-species interaction, it is reasonable to assume that the cross-over diagram will exhibit additional states with the number of bunched atoms at one site smaller than the total number of particles. Also the increase of the intra-species interaction strength $g_{BB}$ and/or correspondingly $N_B$ of the environment might lead to a more complex state, by coupling to the second Bloch band due to an increase of the interspecies interaction energy as a result of a smaller density modulation in the BEC. Furthermore, it is of particular interest how the correlated state will respond dynamically, for example to an external quench through the cross-over diagram. Especially, the possibility of reducing correlations dynamically by lowering the lattice depth is of immediate interest. Beyond that, also dynamically driven setups might exhibit particularly interesting effects, such as persistent currents \cite{ming_pers} induced by the interspecies interaction.

\ack
The authors acknowledge fruitful discussions with J. Schurer, J. Chen and M. Pyzh. P.S. gratefully acknowledges funding by the Deutsche Forschungsgemeinschaft in the framework of the SFB 925 "Light induced dynamics and control of correlated quantum systems". S.K. and P.S. gratefully acknowledge support for this work by the excellence cluster "The Hamburg Centre for Ultrafast Imaging-Structure, Dynamics and Control of Matter at the Atomic Scale" of the Deutsche Forschungsgemeinschaft.

\appendix
\section*{Appendix. Derivation of the effective Hamiltonian}\label{app}
\setcounter{section}{1}
The intra-species interaction among the BEC atoms triggers an excitation of Bogoliubov modes \cite{stringpit,pethsmith}, so that we can write the field operator for the subsystem of the B species as $\hat{\phi}(\text{x})=\phi_0(\text{x}) + \delta \hat{\phi}$ with $\delta \hat{\phi}=\sum_{p} \Big[ u_p(\text{x}) \hat{b}_p+v^{\ast}_p(\text{x}) \hat{b}^{\dagger}_p \Big]$. $\phi_0=\sqrt{N_B}$ is the condensate mode, being spatially homogeneous and depending only on the number of BEC atoms. $u_p(\text{x})$ and $v^{\ast}_p(\text{x})$ are the Bogoliubov modes for a homogeneous BEC and $\hat{b}^{\dagger}_p$ the respective creation operator. For the Hamiltonian of the B species this leads to the well-known effective Hamiltonian within the Bogoliubov approximation $\hat{\bar{H}}_B=\sum_{i} \omega_i \hat{b}^{\dagger}_i \hat{b}_i$, setting $\hbar=L=m_{A/B}=1$.
Plugging this ansatz for $\hat{\phi}(\text{x})$ into the expression of the interspecies coupling Hamiltonian $\hat{H}_{AB}$ we arrive at the following expression

\begin{equation}
\hat{\Delta}_{AB}=g_{AB} \int_{0}^{L} \text{dx} \sum_{i} \Big[ f_i(\text{x}) \hat{b}_i + f^{\ast}_i(\text{x}) \hat{b}^{\dagger}_i \Big] \hat{\chi}^{\dagger}(\text{x}) \hat{\chi}(\text{x}), 
\end{equation}
with $f_i(\text{x})=\phi^{\ast}_0(\text{x}) u_i(\text{x})+\phi_0(\text{x}) v_i(\text{x})$ and $\hat{\chi}^{\dagger}(\text{x})$ the creation field operator for the A species.
Here, we have neglected terms of the order $g_{AB}(\delta \hat{\phi})^{2}$, assuming that $(\delta \hat{\phi})^{2}\ll|\phi_0|^{2} $ and $g_{AB}(\delta \hat{\phi})^{2}\ll \delta \hat{\phi}$. This can be fulfilled for small interspecies interaction strengths and in the case of a small number of Bogoliubov mode excitations. Representing $\hat{\chi}^{\dagger}(\text{x})$ in terms of Bloch states $\chi^{\ast}_k(\text{x})$ with corresponding creation operators $\hat{a}^{\dagger}_k$, the Hamiltonian takes the form

\begin{equation}
\hat{H}\approx\sum_{k} \epsilon_k \hat{a}^{\dagger}_k \hat{a}_k+\sum_{i} \omega_i \hat{b}^{\dagger}_i \hat{b}_i+\sum_{i,k,q} \Big[ g^{kq}_i \hat{b}_i + (g^{qk}_i)^{\ast} \hat{b}^{\dagger}_i \Big] \hat{a}^{\dagger}_k \hat{a}_q, 
\end{equation}
with $g^{kq}_i=g_{AB} \; \langle \chi_k|f_i(\text{x})|\chi_q \rangle$ and $\epsilon_k=\epsilon^{(1)}_k+g_{AB}$ with the single particle energy $\epsilon^{(1)}_k$, corresponding to the $k$-th Bloch state. We omit off-set energy terms in the Hamiltonian stemming solely from the mean-field $\phi_0(\text{x})$. So far, we have rewritten the Hamiltonian using the Bogoliubov approximation. Now, we transform the Hamiltonian in order to decouple the subsystems, using a unitary transformation $\exp(-\hat{S}) \hat{H} \exp(\hat{S})$ with $\hat{S}$ being anti-hermitian. If we neglect terms which are higher than second order in $g_{AB}$, the transformation gives

\begin{equation}
\hat{\widetilde{H}}=\hat{H}_0+\frac{1}{2}[\hat{H}_1,\hat{S}].
\end{equation}
Here, we assumed $[\hat{H}_0, \hat{S}]=-\hat{H}_1$ in order to cancel the two-body term in the Hamiltonian. This condition allows us to determine the operator $\hat{S}$. Making an ansatz, resembling the two-body term $\hat{S}=\sum_{i,k,q} \Big[x^{kq}_i g^{kq}_i \hat{b}_i + y^{kq}_i (g^{qk}_i)^{\ast} \hat{b}^{\dagger}_i \Big] \hat{a}^{\dagger}_k \hat{a}_q$, and determining $x^{kq}_i$ and $y^{kq}_i$ we arrive at the final Hamiltonian 

\begin{equation}
\eqalign{\hat{\bar{H}}=\hat{\bar{H}}_0 + \hat{V}_\text{ind} \quad \text{where,}\\
	\hat{V}_\text{ind}=\frac{1}{2}\sum_{k,q,k^{\prime},q^{\prime}} U_{k,k^{\prime},q,q^{\prime}} \; \hat{a}^{\dagger}_k \hat{a}^{\dagger}_{k^{\prime}}\hat{a}_q\hat{a}_{q^{\prime}}\\
	\hat{\bar{H}}_0=\sum_{k} \epsilon_k \hat{a}^{\dagger}_k \hat{a}_k+\sum_{i} \omega_i \hat{b}^{\dagger}_i \hat{b}_i +\sum_{kq} t_{kq} \hat{a}^{\dagger}_k \hat{a}_q,
}
\end{equation}
with $U_{k,k^{\prime},q,q^{\prime}}=\sum_{i} \Big[ \frac{g^{k^{\prime}q^{\prime}}_i (g^{qk}_i)^{\ast}}{\epsilon_k- \epsilon_q - \omega_i}- \frac{g^{kq}_i (g^{q^{\prime}k^{\prime}}_i)^{\ast}}{\epsilon_k- \epsilon_q + \omega_i} \Big]$ and $t_{kq}=\frac{1}{2} \sum_{r} U_{rkqr}$. The transformation leads to a decoupling of the subsystems, thereby introducing an effective interaction term $\hat{V}_\text{ind}$ for the A species.

\end{document}